\documentclass[aps,prl,superscriptaddress,twocolumn,nopacs,amsmath,showkeys,amssymb,letter]{revtex4}

\usepackage{color}
\usepackage{graphicx}
\usepackage{dcolumn}
\usepackage{bm}

\begin{document}

\title{Direct observation of strain-induced orbital valence band splitting in HfSe$_2$ by sodium intercalation}

\author{T. Eknapakul}
\affiliation {School of Physics, Suranaree University of
Technology, Nakhon Ratchasima, 30000, Thailand}

\author{I. Fongkaew}
\affiliation {School of Physics, Suranaree University of
Technology, Nakhon Ratchasima, 30000, Thailand}

\author{S. Siriroj}
\affiliation {School of Physics, Suranaree University of
Technology, Nakhon Ratchasima, 30000, Thailand}

\author{W. Jindata}
\affiliation {School of Physics, Suranaree University of
Technology, Nakhon Ratchasima, 30000, Thailand}

\author{S. Chaiyachad}
\affiliation {School of Physics, Suranaree University of
Technology, Nakhon Ratchasima, 30000, Thailand}

\author{S.-K. Mo}
\affiliation {Advanced Light Source, Lawrence Berkeley National
Laboratory, Berkeley, CA 94720, USA}

\author{S. Thakur}
\affiliation {Elettra Sincrotrone Trieste, Strada Statale 14 km 163.5, 34149 Trieste, Italy}

\author{L. Petaccia}
\affiliation {Elettra Sincrotrone Trieste, Strada Statale 14 km 163.5, 34149 Trieste, Italy}

\author{H. Takagi}
\affiliation{Department of Physics, University of Tokyo, Hongo, Tokyo 113-0033, Japan} 
\affiliation{Max-Planck Institute for Solid State Research, Heisenbergstrasse 1, D-70569 Stuttgart, Germany}

\author{S. Limpijumnong}
\affiliation {School of Physics, Suranaree University of
Technology, Nakhon Ratchasima, 30000, Thailand} \affiliation
{NANOTEC-SUT Center of Excellence on Advanced Functional
Nanomaterials, Suranaree University of Technology, Nakhon
Ratchasima 30000, Thailand}

\author{W. Meevasana}
\altaffiliation {Corresponding e-mail: worawat@g.sut.ac.th}
\affiliation {School of Physics, Suranaree University of
Technology, Nakhon Ratchasima, 30000, Thailand} \affiliation
{NANOTEC-SUT Center of Excellence on Advanced Functional
Nanomaterials, Suranaree University of Technology, Nakhon
Ratchasima 30000, Thailand}

\date{\today}

\begin{abstract}
	By using angle-resolved photoemission spectroscopy (ARPES), the variation of the electronic structure of HfSe$_2$ has been studied as a function of sodium intercalation. We observe how this drives a band splitting of the p-orbital valence bands and a simultaneous reduction of the indirect band gap by values of up to 400 and 280~meV respectively. Our calculations indicate that such behaviour is driven by the band deformation potential, which is a result of our observed anisotropic strain induced by sodium intercalation. The applied uniaxial strain calculations based on density functional theory (DFT) agree strongly with the experimental ARPES data. These findings should assist in studying the physical relationship between doping and strain, as well as for large-scale two-dimensional straintronics.
\end{abstract}

\maketitle

	Transition metal dichalcogenides (TMDs) have been extensively studied as they exhibit unique physical and electrical properties~\cite{wilson1969,rad2011,opto1,riley2014} with potential applications spanning nanoelectronics, optoelectronics and valleytronics~\cite{nano1,opto1,valley1}. Thus far, most studies have been focused on semiconducting group VIB TMDs (2H-MX$_2$ where M=Mo, W and X=S, Se) which hold enormous potential in creating novel electronic devices such as tuneable optoelectronics and tunnelling field effect transistors~\cite{opto1,lopez,rad2011,rad2,tshen}. Among more than 80 TMD compounds, the so-called 'early TMDs' which adopt a tetragonal phase (1T) are less frequently investigated and hence much remains to be learned from these~\cite{Hfcal1,jfzhao,chho,HfSe2CVD}. For instance, few semiconducting Hf-based TMDs have been reported featuring both a small band gap (ca. 1.13 eV) and a large work function approaching that of Si~\cite{siliconbandgap,Gaizer}. The synthesis of high quality HfSe$_2$ has been achieved recently, e.g. HfSe$_2$ single crystals can be prepared by a chemical vapour transport technique~\cite{HfSe2CVD} and non-distorted HfSe$_2$ thin films can be grown on many different substrates by molecular beam epitaxy, which can be used in the study of TMD heterointerfaces~\cite{hflat2,hfarpes}. These compounds are predicted to be anisotropic with the highest electron mobility amongst TMDs approaching 2,500 $cm^2V^{-1}s^{-1}$~\cite{highmobi}, making it a great candidate for field effect transistors and photovoltaic applications~\cite{Gaizer,highmobi,Kang}.
	
	Alkali metal intercalation achieved by surface electron doping is a powerful method for controlling intrinsic properties in layered-TMDs, due to the weak van der Waals interaction between neighboring chalcogen planes~\cite{Math,Coehoorn,Friend}. Charge accumulation plays an important role in filling the d-orbitals of the transition metal, which results in electronic reconstruction. This leads to novel properties such as the appearance of charge density waves and superconductivity~\cite{jfzhao,Rossnagel}, and quasi-freestanding layers~\cite{quasiMoS2,quasigraphene,quasiSiC}.
	
	The intercalation process affects not only the electronic structure but also the atomic structure, which is strongly coupled with external strain~\cite{distort1,CBdoping}. This relationship is evident from a body of research reporting doping- or strain-related characteristics such as phase transitions and optical phenomena (through e.g. photoluminescence and Raman spectroscopy)~\cite{distort1,straindoping1,dopingphoto}. Computational and experimental studies of strain engineering have significantly advanced in recent decades, exhibiting a wide tunability of optical band gap, electronic structure, and magnetic properties by the introduction of measured degrees of strain~\cite{tshen,straindevice1,strainbandgap,magnetic1,magnetic2}. Such electronic modification is rationalised by lattice deformation, which is occasioned by variation of both strain and doping conditions~\cite{banddeformation1,banddeformation2}. Manipulation of the doping-strain relationship therefore holds promise for sophisticated device engineering, providing a combination of efficient electrolytic gating and lattice straining.
	
\begin{figure}
\includegraphics [width=3.5in, clip]{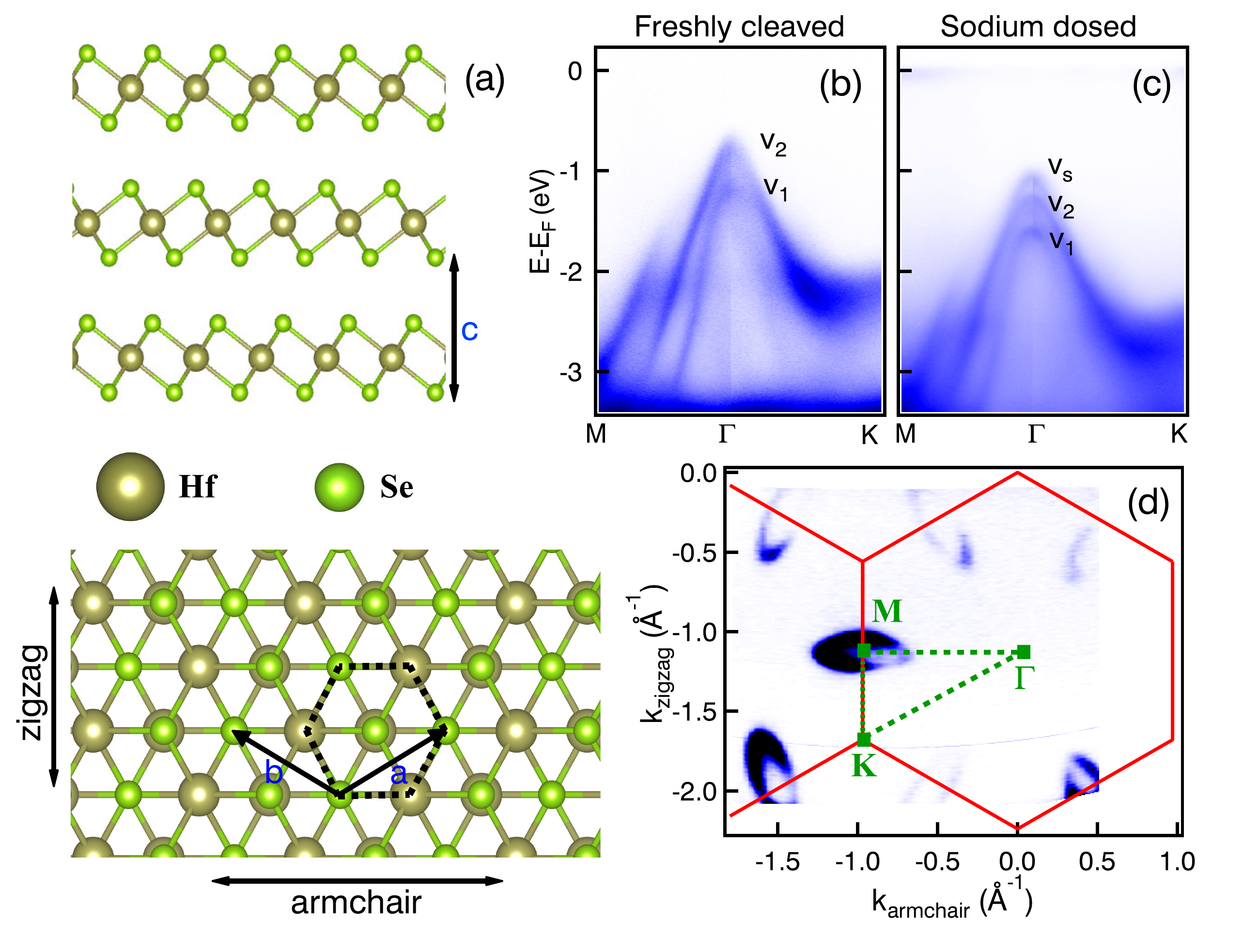}
\caption{\label{FIG1} a) The side and top view atomic structure of 1T-HfSe$_2$. The dashed hexagon represents the in-plane unit cell. Electronic structure of HfSe$_2$ single crystal along M-$\Gamma$-K direction of b) freshly cleaved and c) decent amounts of sodium evaporation, clearly indicates the chemical shift and valence band splitting at the $\Gamma$ point. d) Corresponding Fermi surface map of sodium-dosed sample. }
\end{figure}

\begin{figure*}
\includegraphics [width=6.5in, clip]{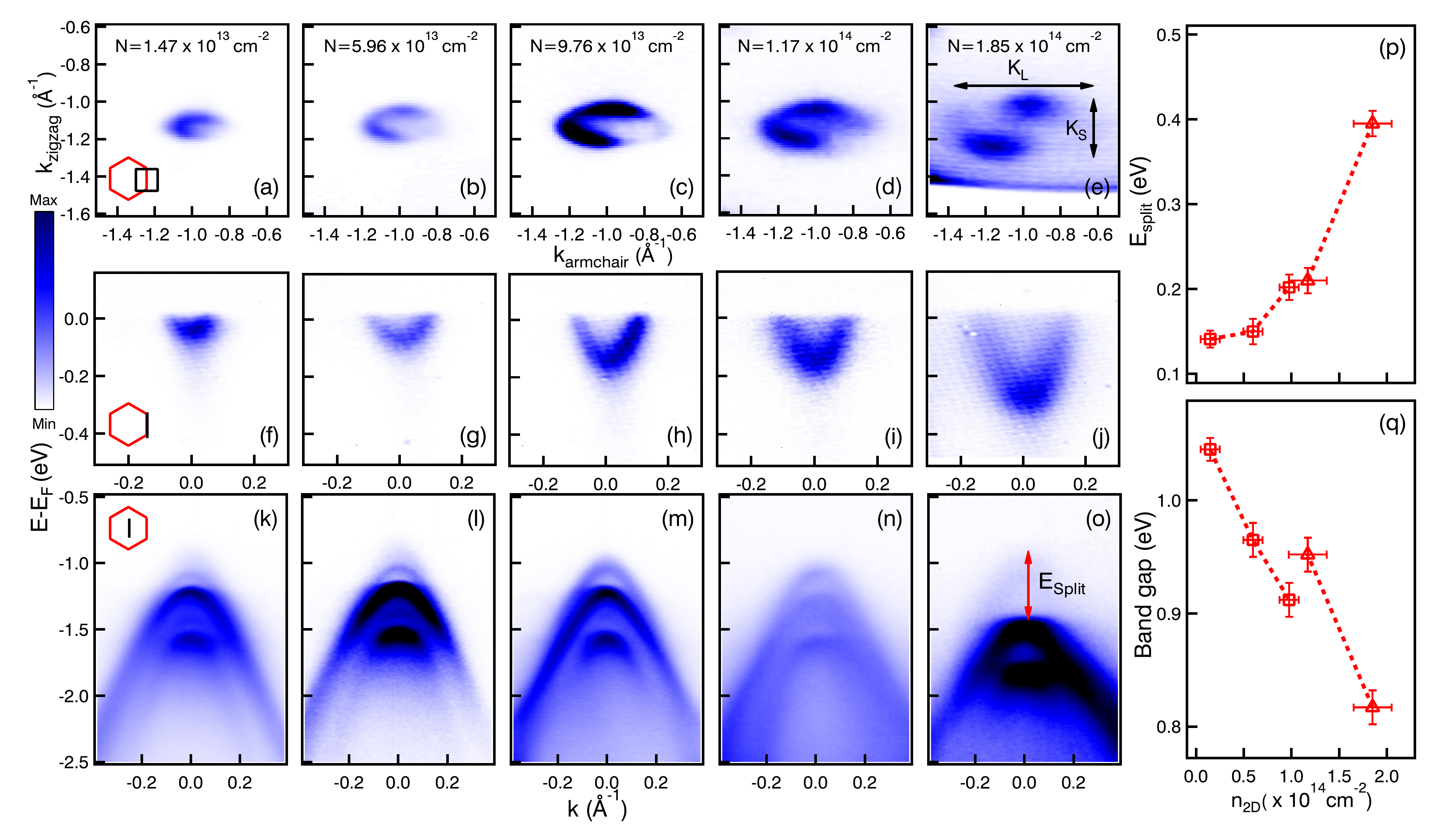}
\caption{\label{FIG2} a)-e) Occupied Fermi surface at M point as a function of electron doping. f)-j) Corresponding two-dimensional Fermi pocket along K-M-K direction. k)-o) Tunability of valence band splitting at $\Gamma$ point corresponding to a)-e). p) Valence band splittings ($E_{split})$ and q) extracted band gap shrinkage  as a function of electron doping. All data were taken at 60 eV photon energy.}
\end{figure*}

Valence band splitting in layered-TMDs by spin-orbit coupling has been proposed to give non-centre splittings as observed in MoS$_2$, MoSe$_2$ and WSe$_2$~\cite{quasiMoS2,Riley_NEC,vbK2,vbK3}. In contrast, calculated valence band splitting at the zone centre can occur by applying external uniaxial strain as a result of deformation potential~\cite{Hfcal1,strainbandsplitting2}. However, direct observation of this phenomenon, a signature of tuneable orbital-valley degree of freedom, has not yet been reported.
In this work, we investigate the influence of sodium intercalation on the electronic structure of HfSe$_2$ by using angle-resolved photoemission spectroscopy (ARPES). First, with increasing doping level, the upward shift of the Fermi level in the conduction band of HfSe$_2$ is observed, as a result of electron donation from the alkali metal into the d-orbitals of the transition metal~\cite{Rossnagel}. Second, a strongly tuneable valence band splitting is observed at the zone centre, accompanied by band gap shrinkage, each of which potentially arise due to the coupling between lattice distortion and uniaxial strain~\cite{distort1,CBdoping}. Our density functional theory (DFT) calculations demonstrate the lattice distortion arising from sodium intercalation and hence the existence of uniaxial strain. The experimentally-observed changes in electronic characteristics agree strongly with the calculated electronic structures under applied uniaxial strain, which therefore supports claims for a strong relationship between doping and strain. These findings are critical in understanding the occurrence of uniaxial strain induced by chemical doping as well as establishing a new synthetic route to chemical gating and strain-engineered devices in layered-TMDs.

	Our 1T-HfSe$_2$ single crystals were grown using a flux growth method and cleaved in ultrahigh vacuum at pressures below 2$\times$10$^{-11}$ torr, providing a pristine surface. The 1T-HfSe$_2$ layered structure resembles CdI$_2$, by featuring Se-Hf in octahedral coordination. Each Se layer within HfSe$_2$ is A-B stacked perpendicular to the c-axis, sandwiched by Hf layers with lattice constants a = 3.71-3.74 $\AA$ and c = 6.14 $\AA$ (see Fig. 1(a))~\cite{hflat1,hflat2}. Our ARPES measurements were performed at BL 10.0.1 of the Advanced Light Source (USA) and BL 10.2R BaDElPh of the Elettra Sincrotrone Trieste (ITA). The photon energies of s- and p- polarised light were set to be 20-80 eV with energy and angular resolution of 10-20 meV and 0.2$^o$, respectively. The sample temperature was set to around 40-80 K. Surface sodium depositions were achieved using a SAES Getter evaporation source.

	Fig. 1(b) and (c) show the ARPES measurement of the HfSe$_2$ electronic structure for the pristine and sodium-evaporated samples. The electronic structure of our pristine HfSe$_2$ along the M-$\Gamma$-K direction is comparable with the bulk, as obtained in previous experiments and calculations~\cite{Hfcal1,hfarpes}. There are no electronic states located at the Fermi level, as expected from the semiconducting bulk (Fig. 1(b)). Moreover, a nondegenerate valence band derived from one unit cell (two monolayers) labelled by v$_1$ and v$_2$ was observed, which can arise from weak interlayer interaction~\cite{hfarpes,vdwcorr1}. After sufficient sodium deposition (Fig. 1(c)), the valence band shifts to a higher binding energy while an additional band (v$_s$) emerged at the $\Gamma$ point, which is a result of electron donation from the alkali metal to the host material~\cite{CBdoping}. The direct population leads to filling of conduction band gives rise to ellipsoidal Fermi contour, indicative of anisotropic behaviour at all M points of the Brillouin zone and consistent with previous calculations (Fig. 1(d))~\cite{Hfcal1,highmobi}.

\begin{figure*}
\includegraphics [width=5.5in, clip]{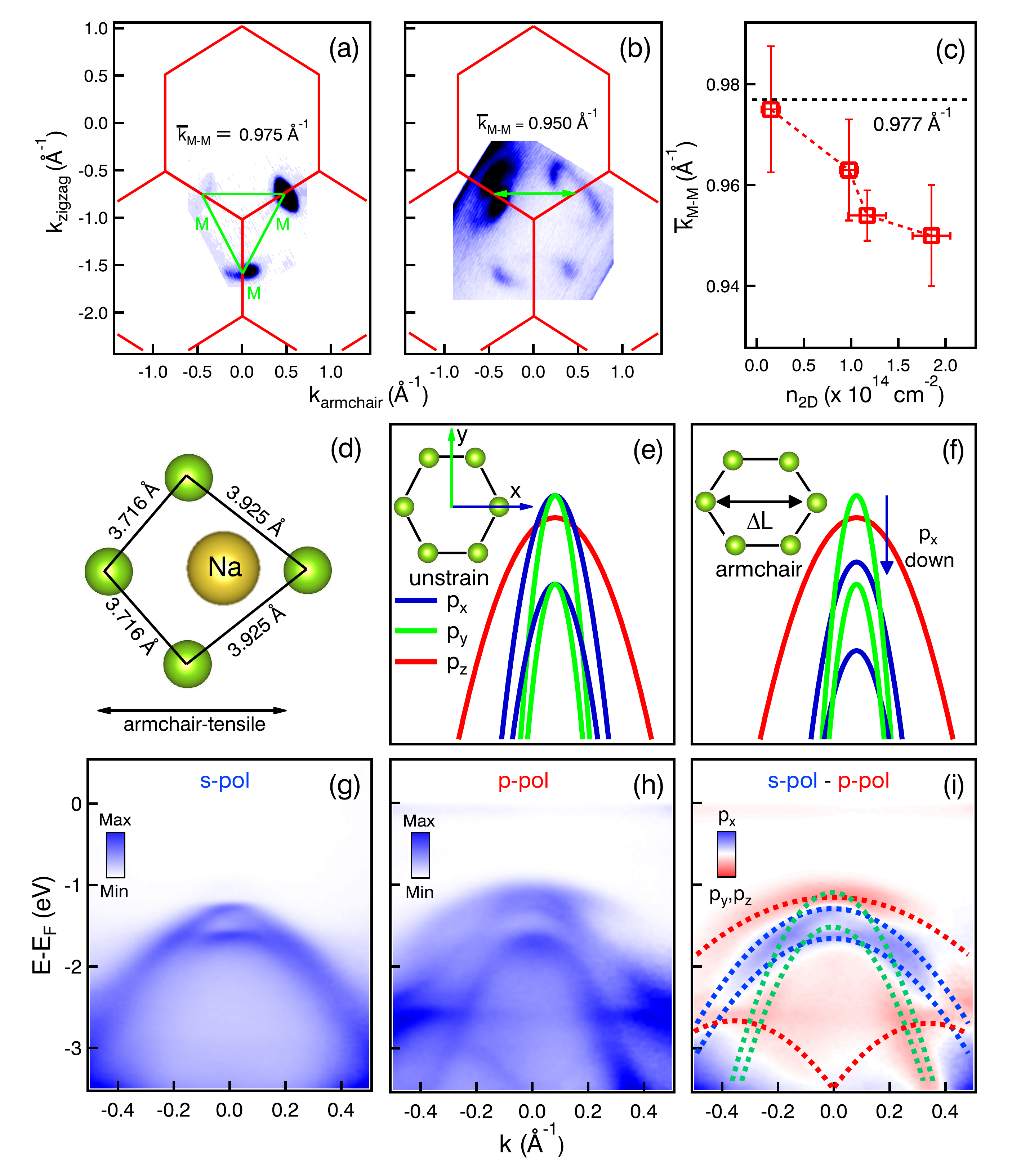}
\caption{\label{FIG3} a) Fermi surface map spanning three M points of a) lightly- and b) heavily sodium-doped surfaces. c) Extracted $\bar{k}_{M-M}$ distance as a function of electron doping. d) Schematic of armchair tensile strain caused by the addition of a sodium atom at the zone centre. e)-f) Schematic band dispersions near valence band maximum indicate the degenerated band as well as the band deformation effect by applying armchair tensile strain. Electronic band dispersion measured along M-$\Gamma$-M direction with g) $s$-polarised and h) $p$-polarised light. i) The difference in spectral weight between $s-$ and $p-$ polarisation reveals the explicit orbital character near the valence band maximum.}
\end{figure*}

We investigated the evolution of the HfSe$_2$ electronic structure as a function of electron doping, achieved by increasing exposure time for sodium evaporation (with constant current (I = 6.2 A)). As shown in Fig. 2(a)-2(e), by increasing exposure time, the surface carrier density, as determined by its Luttinger area (N$_{2D}$~=~$g_v\frac{k_Sk_L}{2\pi}$, where g$_v$~=~3 is valley multiplicity for all occupied ellipsoidal pockets, k$_S$ and k$_L$ are short and long ellipsoidal radii), increases with increasing time for the same sample. The maximum surface carrier density can reach $1.85\times10^{14}~cm^{-2}$ or approximately 0.27 electrons per surface unit cell higher than those previously reported~\cite{Riley_NEC,CBdoping}. This suggests that HfSe$_2$ has a high capacity for extrinsic dopant atoms via its van der Waals (vdW) layer, making it suitable as a host material or substrate. We found a clear increase in the valence band splitting ($E_{split}$~=~v$_s$-v$_2$) at the valence band maximum, found by fitting the energy distribution curves (EDCs) around the $\Gamma$ point of two distinct valence bands up to 400~meV (Fig. 2(k)-2(o)), as summarised in Fig. 2(p)). By extracting the gap between the conduction band minimum (M point, Fig. 2 (f)-2(j)) and valence band maximum ($\Gamma$ point, Fig. 2(k)-2(o)), we found a continuous decrease in the indirect band gap up to a maximum splitting of 280~meV (Fig. 2 (q)). This observation could be ascribed to surface accumulation or intercalation caused by the introduction of the alkali metal at the surface.

	Unlike previous reports of alkali deposition on several TMDs~\cite{Riley_NEC, electricfield}, our observed band splitting can not be explained by the introduction of a surface electric field from charge accumulation. For example, the existence of valence band splitting at K point can be obtained by applying an electric field at the WSe$_2$ surface, but could not be seen at the $\Gamma$ point of HfSe$_2$. This may be the critical difference between 1T and 2H symmetry (see the difference in Supplemental Material). In other words, the intercalation should be the dominant effect in HfSe$_2$ system and a rationale is needed to include strain here. 
	
	Based on the experimental findings, we then calculated the band structure of HfSe$_2$ under different strain conditions induced by sodium intercalation. Calculations were carried out within the framework of density functional theory with projector augmented wave potentials (PAW)~\cite{cal1} as implemented in the VASP code. The PBE approximation is used for the exchange correlation terms~\cite{cal2,cal3} with van der Waals corrections carried out using the DFT+D3 method~\cite{vdwcorr1}. An 11 $\times$ 11 $\times$ 1 and 5 $\times$ 5 $\times$ 1 Monkhorst-Pack~k-point grid were used for bulk and supercell calculations, respectively~\cite{Monkhorst}. The electron wave functions were described using a plane wave basis set with the energy cutoff at 520 eV.

	In addition, the Fermi surface maps spanning three M points of the Brillouin zone of lightly- and heavily-doped samples are shown in Fig. 3(a) and 3(b), respectively. By using $\bar{k}_{M-M}$ distance of 0.977 $\AA^{-1}$ (as given by $\frac{2\pi}{\sqrt3a}$) for a fresh sample to be used as a reference, we found Brillouin zone shrinkage up to 2.76~$\%$, as determined by a decreasing of $\bar{k}_{M-M}$ as a function of electron doping (Fig. 3(c)). This could be rationalised by the intercalation of sodium into HfSe$_2$ layers, hence enlarging the unit cell in real space~\cite{enlarge1,enlarge2}. Moreover, the Brillouin zone shrinkage appears larger in the reciprocal tensile direction than the zigzag direction, suggesting a preference for expansion in the armchair direction in real space. Fig. 3(d) represents the possibility of introducing armchair tensile strain when adding sodium at unit cell centre. The fully relaxed atomic structure of Na$_{0.25}$HfSe$_2$, illustrating the lattice distortion and expansion effects arising from the introduction of a single sodium atom into the centre of a 2 $\times$ 2 $\times$ 2 cell, are shown in the Supplemental Material. 
	
	The calculated electronic band structure of anisotropic lattice distortion in HfSe$_2$ is supported by the experimental ARPES results, where three different hole-like bands are observed near the valence band maximum of HfSe$_2$ arising from Se $p_x,p_y, and~p_z$ orbitals~\cite{Hfcal1}. The most heavily 'hole-like' band is $p_z$, followed by $p_x$ then $p_y$. Fig. 3(e) and 3(f) illustrate schematic band deformation arising from external uniaxial strain. In HfSe$_2$, the top of the valence band is contributed to by Se p-orbitals, where p$_x$ and p$_y$ orbitals are degenerate at the valence band maximum, while the p$_z$ orbital is located at slightly higher binding energy (Fig. 3(e)). The relative positions of the orbitals can be rearranged by applying uniaxial strain. As shown in Fig. 3(f), the p$_x$ orbital is shifted to higher binding energy resulting in the contribution of p$_y$ to the valence band maximum under armchair tensile strain conditions. Light polarisation-dependent measurements are best suited to reconciling orbital ordering with those bands expected within our geometry~\cite{Lewis_pol}. Fig. 3(g) shows the electronic band structure measured by s-pol where the most intense bands are $p_x$ orbitals. In contrast, p-pol measurement reveals more orbitals than s-pol, where $p_y$ and $p_z$ appear most notable (Fig. 3(h)). Fig. 3(i) shows the intensity difference obtained by subtracting s- and p- pol data, clearly indicating the valence band splitting of $p_x$ and $p_y$ orbitals in our doped sample. Overall, it is shown that sodium evaporation/intercalation can lead to lattice distortion. The distorted HfSe$_2$ prefers to rearrange in armchair tensile strain~\cite{ZnO,Ribb2}, as confirmed by our polarisation-dependent ARPES data (Fig. 3(i)).
	
\begin{figure}
\includegraphics [width=3.5in, clip]{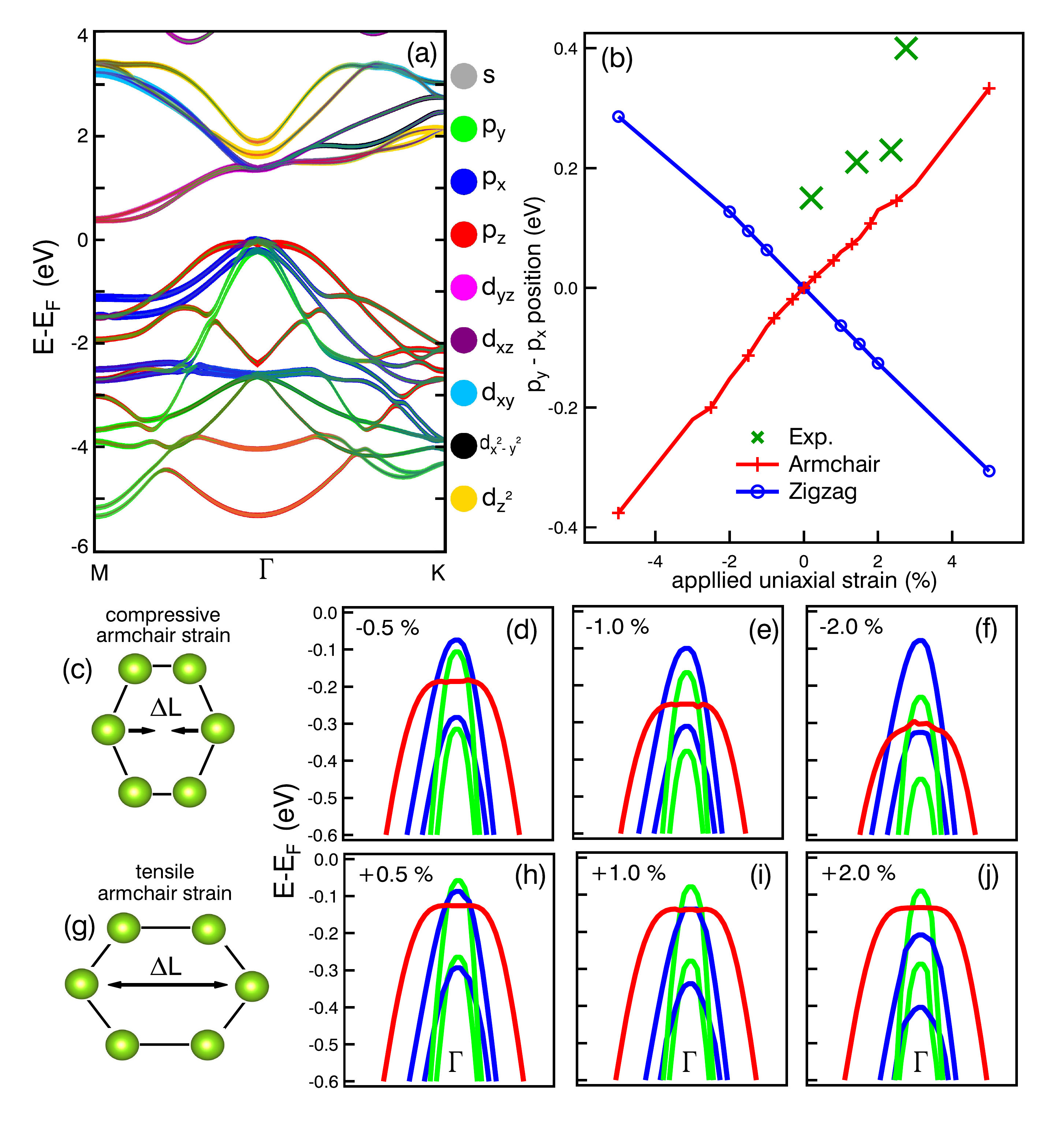}
\caption{\label{FIG4} a) The calculated orbital projected band structure of HfSe$_2$ using DFT calculation+vdW correction. b) Extracted valence band splitting ($p_y-p_x$), corresponding to ($v_s-v_2$) of the ARPES data as a function of applied uniaxial strain. Dominant electronic structure at the top of the valence band by varying between c)-f) compressive and g)-j) tensile armchair-uniaxial strains.}
\end{figure}

The ARPES results suggest that sodium atoms can occupy a maximum of 27$\%$ of surface unit cells, which is comparable to the computational Na$_{0.25}$HfSe$_2$ model. The increase in the valence band splitting brought on by sodium doping is a result of an increased uniaxial strain in the doped HfSe$_2$ layer. Electronic structure calculations for HfSe$_2$ have been performed under armchair and zigzag uniaxial strains ranging from -5$\%$ to 5$\%$. The unstrained orbital-projected electronic structure of HfSe$_2$ is shown in Fig. 4(a) representing Se p-orbitals located at the top of valence band. Fig. 4(d)-4(j) represent the zoomed-in electronic structures around the $\Gamma$ point between -2 to +2 $\%$ uniaxial armchair strain, where minus and plus signs indicate compressive (Fig. 4(c)) and tensile (Fig. 4(g)) strain respectively. The band deformation plays an important role in varying relative band positions (and therefore band gap) and valence band splittings in these calculations. The p$_x$(p$_y$) orbitals are shifted in energy monotonically while p$_y$(p$_x$) orbitals are essentially constant as a function of applied tensile(compressive) strain, resulting in opposite sign of valence band splitting. Regarding the two previously-mentioned in-plane p-orbital bands, which are degenerate at the valence band maximum for bulk HfSe$_2$, applying uniaxial strain significantly lifts the degeneracy and therein changes the electronic band gap. The calculated valence band splitting (p$_y$~-~p$_x$) reaches 333 and -376 meV in 5\% tensile and compressive armchair strain, respectively. The zigzag uniaxial strain has also been performed, yielding valence band splitting in the opposite direction (Fig. 4(b)). Larger valence band splittings obtained from the ARPES data (green marks in Fig. 4(b) were found as compared with the calculated tensile strain values. This may be describable by anisotropic lattice distortion in which the asymmetrically-enlarged lattice (tensile armchair strain) is compressed in the zigzag direction to maintain its volume. Overall however, we note that the observed orbital valence band splitting are generally consistent with armchair-tensile strain.
		
		In summary, we have observed an increase in orbital valence band splitting, band gap shrinkage and anisotropic unit cell enlargement by evaporating sodium onto HfSe$_2$ surfaces corresponding to 27\% surface doping concentration (equivalent to 0.27 electrons donated per surface unit cell). Our experimental ARPES data are well-supported by computational DFT calculations of fully relaxed Na$_{0.25}$HfSe$_2$, where the calculations showed the HfSe$_2$ layer anisotropically distorted with about 5$\%$ strain due to sodium intercalants. Overall, it can be concluded that the presence of effective armchair-tensile strain is induced by sodium intercalation. We have also demonstrated the possibility of orbital control at the valence band maximum with 'valley physics'. In future work, this could be achieved by applying opposite uniaxial strain by introduction of different cationic (Na, K) or anionic (Cl, Br) intercalants or ionic liquids. These observations support an overall understanding of doping/strain physics to facilitate exploration of novel phenomena among layered-TMDs. This could also be used as a direct technique for large-scale strain engineering in tuning optical and electronic properties as well as for straintronics such as nanoscale stress sensors and tuneable photonic devices~\cite{straintronics,apps}.

\begin{acknowledgments}
We acknowledge P. D. C. King, P. Songsiriritthigul and T. Kongnok for supplying helpful discussion. This work was supported by Thailand Research Fund (TRF) and Suranaree University of Technology (SUT) (Grant No. BRG5880010) and Research Fund for DPST Graduate with First Placement (Grant No. 021/2555). The computing resources were provided by Theoretical Computation Section, SLRI, Thailand. SS acknowledges the Royal Golden Jubilee Ph.D. Program (Grant no. PHD/0007/2555). We thank ALS and Elettra for the allocation of synchrotron radiation beamtimes. The Advanced Light Source is supported by the U.S. Department of Energy, Office of Science, User Facility under contract No. DE-AC02-05CH11231.
\end{acknowledgments}

\end{document}